\documentclass[prx, twocolumn, amsmath,amssymb,superscriptaddress]{revtex4}
\usepackage{graphicx}
\usepackage{amssymb}
\usepackage{epstopdf}
\usepackage{bbold}
\usepackage{xcolor}
\usepackage{makecell}
\usepackage{mdframed}
\usepackage{float}
\usepackage{textgreek}

\begin{document}
\title{Tissue hydraulics: physics of lumen formation and interaction}
\author{Alejandro Torres-S\'anchez}
\thanks{Equal contribution}
\affiliation{The Francis Crick Institute, 1 Midland Road, NW1 1AT, United Kingdom}
\author{Max Kerr Winter}
\thanks{Equal contribution}
\affiliation{The Francis Crick Institute, 1 Midland Road, NW1 1AT, United Kingdom}
\author{Guillaume Salbreux}
\thanks{To whom correspondence should be addressed: guillaume.salbreux@unige.ch}
\affiliation{The Francis Crick Institute, 1 Midland Road, NW1 1AT, United Kingdom}
\affiliation{University of Geneva, Quai Ernest Ansermet 30, 1205 Gen\`eve, Switzerland}
\date{\today}

\begin{abstract}
Lumen formation plays an essential role in the morphogenesis of tissues during development. Here we review the physical principles that play a role in the growth and coarsening of lumens. Solute pumping by the cell, hydraulic flows driven by differences of osmotic and hydrostatic pressures, balance of forces between extracellular fluids and cell-generated cytoskeletal forces, and electro-osmotic effects have been implicated in determining the dynamics and steady-state of lumens. We use the framework of linear irreversible thermodynamics  to discuss the relevant force, time and length scales involved in these processes. We focus on order of magnitude estimates of physical parameters controlling lumen formation and coarsening.
\end{abstract}
\maketitle

Lumen formation is ubiquitous in developmental biology \cite{Datta2011, Sigurbjornsdottir2014, Navis2015, blasky2015polarized, navis2016pulling}. The successful opening of fluid filled spaces within tissues is crucial for a zygote to grow into a topologically complex adult organism containing multiple cavities and networks of tubes. There are many prominent examples of lumen formation in development \cite{Navis2015}, and before turning to the physics of lumen formation and their interactions, we first discuss here some of these examples, illustrating the variety of roles lumen formation plays by contributing to morphogenesis, symmetry breaking, cell fate specification, or size control.

Lumen expansion driven by fluid flow, which is the essential process we discuss here, is one the fundamental mechanisms by which lumens can form \cite{Navis2015}. We review the physics of this process in more detail in section \ref{sec:single_lumen}. Fluid accumulation has been implicated or proposed to participate e.g.~in zebrafish gut formation \cite{Bagnat2007}, brain ventricular expansion {\cite{lowery2005initial, zhang2010establishment}}, otic vesicle formation {\cite{hoijman2015mitotic,Swinburne2018,Mosaliganti2019}}, formation of Kuppfer's vesicle \cite{dasgupta2018cell} or in the developing mouse salivary gland \cite{nedvetsky2014parasympathetic}.
During embryogenesis, fluid pumping by polarised cells in the epiblast, rather than cell apoptosis, acts as a lumen formation mechanism in the mouse pro-amniotic cavity {\cite{Bedzhov2014, Kim2021-yy}}. 
The prospective pro-amniotic cavity forms in a similar way in the epiblast of human embryos, where a lumen opens in the centre of a rosette of polarised cells, and in the absence of apoptosis \cite{Shahbazi2016}.

\begin{figure}[h!]
	\centering
	\includegraphics[width=0.9\linewidth]{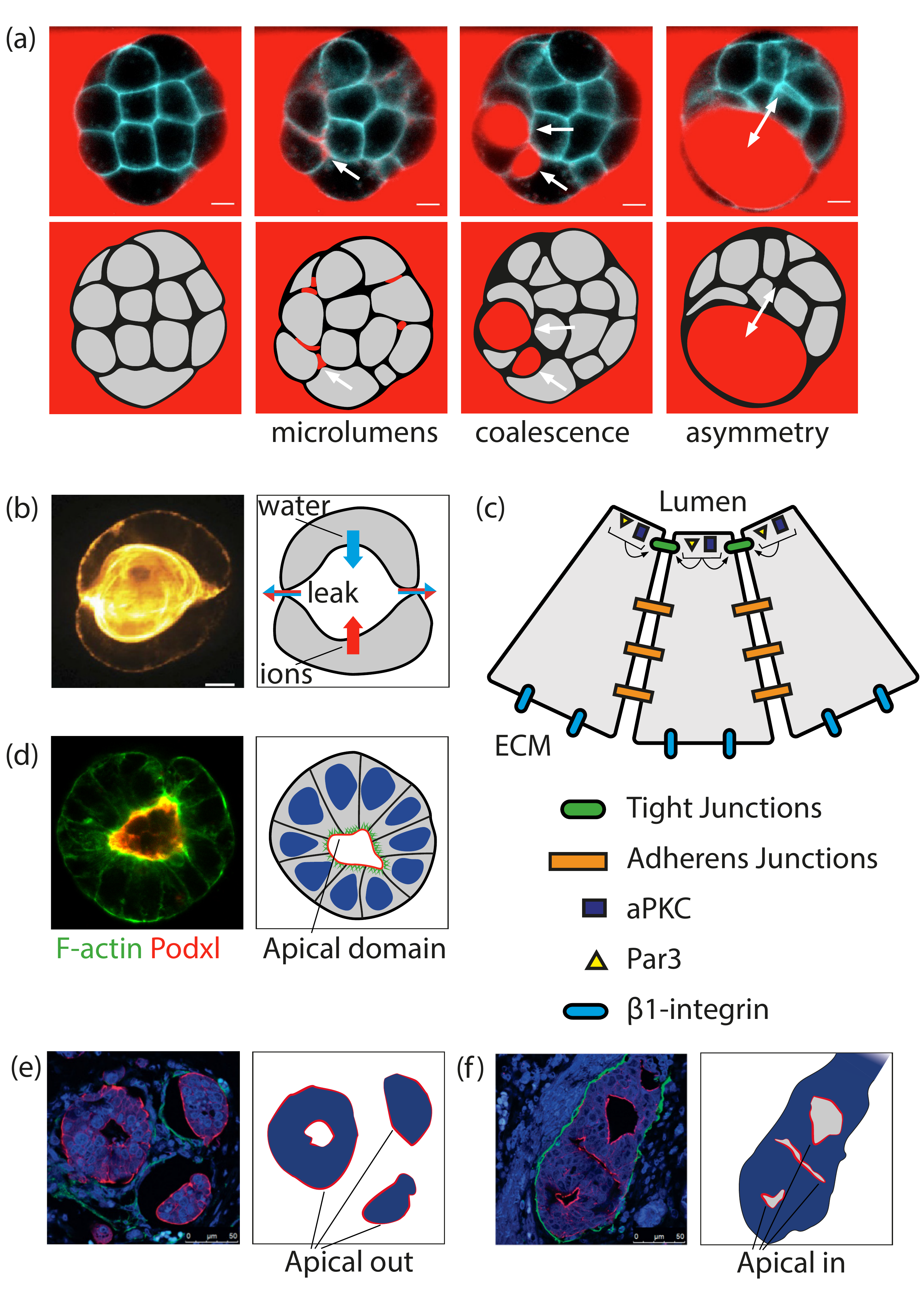}
	\caption{Examples of lumen formation. (a) Formation of the mouse blastocyst. The surrounding fluid is pumped into the centre of the embryo, forming microlumens. These microlumens then coalesce, and eventually form one large lumen, the blastocoel. Top images: cyan, plasma membrane label; red, Dextran; Scale bar, $10$ \textmu m \cite{Dumortier2019}.  (b) A lumen formed between two rat hepatocytes. This process has been modelled as a balance between active pumping into the cavity, and paracellular leakage out of it \cite{Dasgupta2018}.  {Scale bar, 2 \textmu m. } (c) A schematic of the polarised structure of epithelial cells surrounding a lumen. The basal domain is established by sensing the external environment via $\beta$1-integrins \cite{Manninen2015}. Tight junctions form on the apicolateral interfaces. The apical domain is characterised by apical proteins such as aPKC and PAR3 {which then relocalise to tight junctions \cite{Datta2011}}. {Opposite cell polarity orientation with respect to the lumen position can also occur \cite{wang1990steps}.}  (d) A spherical aggregate of mouse epiblast stem cells. The anti-adhesive molecule podocalyxin (Podxl) is expressed at the apical domain. Aggregates of epiblast cells recapitulate lumen forming events in the mouse embryo (image courtesy of M. Shahbazi). (e), (f) Colorectal cancer samples demonstrating apical polarity directed outward (e) and inward (f); {green: D2-40 (lymphatic endothelial cells); red: LMO7 (apical membrane of cancer cells); blue: DAPI} \cite{Okuyama2016}. All images are reproduced with permission.}
	\label{Fig:introduction}
\end{figure}

\textbf{Lumen coarsening. }
Larger lumens form in some cases through the coarsening of smaller lumens; we discuss this process in more detail in section \ref{sec:lumen_interactions}.  In the zebrafish gut, multiple small lumens open via fluid accumulation driven by paracellular {and transcellular} ion transport \cite{Bagnat2007, Alvers2014}. These lumens subsequently fuse by tissue remodelling where cell-cell adhesions are lost in the tissue bridges separating neighbouring lumens. 
Fusion is a distinct process to lumen nucleation, as demonstrated by \textit{smoothened} mutants that exhibit small lumens which fail to fuse \cite{Bagnat2007, Alvers2014}. {Similarly in early lumen formation in zebrafish inner ear morphogenesis, two small, initially unconnected lumens appear to coalesce and fuse \cite{hoijman2015mitotic}.}
A coarsening mechanism has also been reported during the formation of the {blastocoel}, the first lumen forming event in both mouse and human development {\cite{Dumortier2019,Ryan2019,Zernicka-Goetz2005,Frankenberg2016, maitre2017mechanics}}. The successful formation of the blastocyst is crucial for the viability of the embryo \cite{Marikawa2012, Kim2021-yy}, and occurs by the active pumping of fluid into the centre of the embryo by an outer layer of polarised cells. Initially, a large number of microlumens form in the intercellular space by hydraulic fracturing of cell-cell adhesions. Then, these combine into a single, large lumen (see Fig.~\ref{Fig:introduction}a, where Dextran fluid labelling allows to visualise the transient fluid accumulation at cell-cell contacts \cite{Dumortier2019}). {Differences of cell contractility within the embryo are thought to provide directionality to this coarsening process and guide the final position of the blastocoel \cite{Dumortier2019}.}

\textbf{Interplay between lumen growth and epithelial mechanics. }
 The interplay of luminal pressure and the rupture and healing dynamics of the surrounding epithelium has been proposed to act as a size control mechanism. In the mouse blastocyst, increased luminal pressure leads to increased tension in the trophectoderm, the epithelium lining the blastocoel. Above a critical tension, cell-cell adhesions cannot be maintained during mitosis, resulting in the temporary rupture of the blastocyst. This mechanism results in the cavity radius oscillating about some average value \cite{Chan2019}, following a mechanism which had also been proposed to explain tissue oscillations observed during {\it Hydra} regeneration \cite{futterer2003morphogenetic,Kucken2008}.  These oscillations have been modelled by considering a spherical elastic shell surrounding a pressurised lumen, which ruptures {and forms a tear} above a critical surface tension \cite{Ruiz-Herrero2017}. Following rupture, luminal fluid flows out of the tear, the tension decreases, and the tear heals at some lower tension. Denoting $\sigma_c$ and $\sigma_h$ the surface tensions at which the tear opens and closes, $E$ the 2D elastic modulus of the epithelium, and assuming $|\sigma_c-\sigma_h|\ll E$, this process results in oscillations around a mean radius value for the lumen $R\sim R_0(1+ \sigma_c/E)$ with $R_0$ the radius {of the tensionless shell}, and amplitude $\Delta R\sim (\sigma_c-\sigma_h)/E$ \cite{Ruiz-Herrero2017}. 
 
Luminal pressure has also been studied, and directly measured, in the developing zebrafish inner ear \cite{Swinburne2018,Mosaliganti2019}. Here it has been proposed that hydrostatic pressure arising from stresses developing in the epithelium can act as a correcting size mechanism: indeed a decrease in size of the otic vesicle results in stress relaxation in the epithelium, lowering the hydrostatic pressure in the lumen and leading to faster growth \cite{Mosaliganti2019}. 

On a smaller scale, lumen expansion has been associated during zebrafish sprouting angiogenesis with pressure-driven inverse membrane blebbing, local membrane protrusions resulting from membrane/cortex detachment \cite{gebala2016blood}.

\begin{figure*}
\begin{mdframed}
{\bf Interaction of charged surfaces.}
 Here we discuss the physics of the interaction of charged surfaces in an electrolyte solution. We consider an electrolyte solution with 1:1 charge ratio and valence $z$ (e.g. NaCl, with $z=1$). The steady-state concentration profile of ionic species is described for small electric potential by the linearised Poisson-Boltzmann equation \cite{butt2013physics},
\begin{align}
\nabla^2 \phi = \frac{2 \bar{c} z^2 e^2}{\epsilon kT} \phi = \frac{1}{\ell^2_{\rm DH}} \phi,
\end{align}
where $\epsilon$ is the dielectric permitivity of the solvent,  $e$ the charge of an electron, $\bar{c}$ the background concentration of electrolyte, $k$ the Boltzmann constant, $T$ the temperature, and $\phi$ the electric potential. We have also introduced the Debye-H{\"u}ckel screening length $\ell_{\rm DH} = \sqrt{\epsilon k T/(2 \bar{c} z^2 e^2)}$, which determines the length scale on which electrostatic interactions occur in an electrolyte solution. Substituting in a typical biological concentration of ${\rm Na}^+$, ${\rm Cl}^-$ $\bar{c} \sim 100$ mM \cite{butt2013physics}, $z=1$, $e=1.6\times10^{-19}$ C, $\epsilon=\epsilon_r \epsilon_0 \approx 74.5 \epsilon_0 \approx 7\times10^{-10}\,\text{F/m}$ with $\epsilon_0$ the vacuum permittivity and $\epsilon_r$ the relative permittivity of water \cite{butt2013physics}, and $kT=4\times10^{-21}$ J, we get $\ell_{\rm DH} \approx 1$ nm. For comparison, the typical length of a E-cadherin junction is 25 nm \cite{Perez2004}. Contact repulsion through negative charges would therefore have to occur on very close contact between cellular interfaces.

How strong can the force from such a repulsive contact be? The pressure acting on two charged planar interfaces separated by a distance $d$ and carrying a charge per unit area $e n$ is, for small electric potentials \cite{butt2013physics}:
\begin{align}
\label{eq:electrostatic_interaction_pressure}
P = \frac{2e ^2 n^2}{\epsilon} \exp\left(-\frac{d}{\ell_{\rm DH}}\right).
\end{align}
The rupture force of a single cadherin bond is of the order of $100$pN \cite{panorchan2006single}. Approximately 400 cadherin dimers span a cell-cell interface of area $100$\textmu m$^2$ (taking a cadherin density of $80$ molecules/\textmu m$^2$, with a ratio of monomers to dimers of 17:1 \cite{Katsamba2009}), giving a rupture pressure of $\sim 400$Pa.
According to Eq.~\eqref{eq:electrostatic_interaction_pressure}, the charge density required to generate a repulsive pressure leading to cadherin rupture is $2\times 10^3$ \textmu m$^{- 2} <n<6\times 10^8 $\textmu m$^{- 2}$ for distances $d$ ranging between 0 and $25$nm. The charge of podocalyxin is approximately -16e {at neutral pH}, and is supplemented by approximately 20 sialic acid molecules, each of which carries a charge of $-e$ \cite{proteincalcwebsite,Kerjaschki1985,Varki2009}. Consequently, these charge densities correspond to between {$60$ and $2\times 10^7$} podocalyxin molecules {per \textmu m$^2$}. A lower bound of podocalyxin density on rat epithelial cells of {200}\textmu m$^{-2}$ has been reported \cite{Kerjaschki1985}. We conclude that repulsive electrostatic forces could be high enough to separate cell membranes, but the corresponding forces can only act at very short distance. This analysis is however highly simplified, as notably the size of charged membrane proteins is not small compared to the Debye length.
 \end{mdframed}
\end{figure*}

\textbf{Lumen opening via electrostatic interactions.}
In the developing mouse aorta, a vascular lumen is initiated at cellular interfaces between endothelial cells \cite{strilic2009molecular}. It has been suggested that repulsion of negatively charged tissue interfaces contribute to this lumen formation process \cite{Strilic2010}.  Such repulsive interactions are thought to be due to charged ``anti-adhesin'' molecules at the cell surface, such as podocalyxin, which can be secreted into the lumen nucleation site, with electrostatic forces separating adhering cells. Podocalyxin is a transmembrane protein and a constituent of the glycocalix \cite{nielsen2008novel} whose ``anti-adhesive'' effect has been documented e.g. through its effect in decreasing cell aggregation in MDCK cells and has been shown to be dependent on sialic acid  \cite{Takeda2000}. Since electric charges are screened on short distances of $\sim $nm  in electrolyte solutions, physical arguments indicate that such electrostatic repulsive interactions occur if cells, or their glycocalyx layer, are in very close contact. We give in the box ``Interaction of charged surfaces'' order of magnitude estimates of the interaction of charged plates to illustrate this discussion.  {If such a close contact occurs, one may expect the arrangement of mucins on the cell surface, which evokes a polyelectrolyte brush \cite{kuo2018physical}, to give rise to cellular interfaces interactions  which correspond to the disjoining pressure of grafted polyelectrolyte on surfaces \cite{pincus1991colloid}}. Podocalyxin might also have a more indirect effect on cell adhesion, as it has also been shown to promote the formation of microvilli at the cell surface \cite{nielsen2007cd34}.

\textbf{Lumen contribution to tissue patterning.}
 As well as physically separating groups of cells, lumens contribute to patterning by the presence of signalling molecules in the luminal fluid, accumulating in microlumens \cite{durdu2014luminal}. These microlumens can act as signalling hubs, as secreted diffusible molecules can easily reach cells enclosing the lumen, ensuring their coordinated response. During mouse blastocyst formation, the specification and segregation of the epiblast and primitive endoderm, with the latter being eventually in contact with the lumen, occurs concomitantly with lumen expansion and is impaired if lumen expansion does not proceed normally \cite{Ryan2019}.
 
 {Fluid exchange between cells and the extracellular medium has recently been shown to play a role in cell fate specification during oogenesis in \textit{C. elegans}. Here, an hydraulic instability amplifies volume differences in germ cells, resulting in the establishment of a heterogeneous distribution of germ cell volumes. Smaller cells within this population tend to undergo apoptosis, thus providing an example of mechanically induced cell fate specification \cite{Chartier2021}.}
 
\textbf{Lumen formation in vitro.}
{\it In vitro} systems allow the mechanisms of lumen formation to be studied in isolation. An example of such an isolated lumen forming between two rat hepatocytes is shown in Fig.~\ref{Fig:introduction}b. This process of lumen formation has been described theoretically with a model involving the balance of fluid pumping by the cells into the lumen with paracellular leakage out of it \cite{Dasgupta2018}. Hepatocytes have also been used to study the formation of anisotropic lumens which do not assume a spherical shape and can have a biased position along cellular interfaces \cite{Li2016}. 

Madin-Darby Canine Kidney (MDCK) cells are a common {\it in vitro} system for studying lumen formation, as they readily self-organise into polarised epithelial spheres when grown in 3D \cite{McAteer1986} and can form with opposite polarities when grown in suspension ({basal side facing the lumen}) or in collagen gels \cite{wang1990steps}. In cysts formed in collagen gels, the apical side is facing the lumen, and apicobasal polarity in MDCK cells is initiated by the sensing of collagen in the extra-cellular matrix (ECM) via $\beta$1-integrins \cite{Yu2005}. This signal induces reorganisation of the cytoskeleton, and the accumulation of apical polarity proteins, such as aPKC and Par3, at a region of the inward facing cell membrane called the apical membrane initiation site \cite{Yu2008,Bryant2010, Datta2011, Sigurbjornsdottir2014}. This is shown schematically in Fig.~\ref{Fig:introduction}c. Having established an inward facing apical domain, the lumen begins to form by active pumping.  
As well as in spherical cysts, the mechanics and hydraulics of MDCK cells have long been studied in doming epithelial monolayers, which form local blisters by detachment from the substrate \cite{Cereijido1981-bz,Tanner1983-fi,Fernandez-Castelo1985-pi}. Blisters are hydraulically connected through the lateral intercellular space (LIS), which is {weakly} compliant to hydrostatic pressure, or through flows occurring basally \cite{Timbs1996-gt}. When the epithelium is subjected to a sudden increase of basal hydrostatic pressure \cite{Casares2015}, hydraulic fractures appear in the LIS which are reminiscent of the microlumens forming in the mouse embryo \cite{Dumortier2019}. Recently, epithelial blisters have been employed to demonstrate the ability of the epithelium to sustain large deformations at constant tension, a hallmark of superelasticity \cite{Latorre2018}.

Recent experiments have demonstrated the extent to which embryonic stem cells can self-organise \cite{Harrison2017, Sozen2018, Sozen2019}. Aggregates of a single stem cell lineage, epiblast cells, have been studied {\it in vitro}, with a particular emphasis on their ability to form lumens \cite{Bedzhov2014} (Fig.~\ref{Fig:introduction}d). Similar to MDCK aggregates, an internal apical domain is established through a mechanism which is dependent on $\beta$1-integrin signaling \cite{Bedzhov2014, Mole2021}. In both mouse and human cells, successful lumen formation is coordinated by the transition from naive to primed pluripotent states. As cells exit naive pluripotency, {an Oct4-governed transcriptional program} results in the expression of podocalyxin, and ultimately lumen nucleation \cite{Shahbazi2017}.


{In intestinal organoids, several studies have found a coupling between lumen dynamics and morphogensis. Lumen volume decrease driven by enterocytes, one of the intestinal organoid cell populations, contribute to bulging deformations of initially spherical organoids, possibly because luminal volume reduction results in increased compressive stresses in the epithelium \cite{Yang2020}. A recent work found that stem cell zones, regions of the organoid containing intestinal stem cells, could undergo fissions and deformation events which are dependent on phases of luminal expansion and collapse \cite{Tallapragada2021}. Merging of multiple intestinal organoids have also been shown to result in the formation of {\it in vitro} intestinal tubes which preserve a continuous lumen surrounded by a monolayer epithelium \cite{Sachs2017}.}

\textbf{Lumens in disease.}
Outside of developmental biology, the formation, maintenance or disruption of lumens play a role in {e.g.~}polycystic kidney disease and cancer. In polycystic kidney disease, the renal epithelial tubules are abnormally enlarged, and ions pumps and transporters are mislocalized at the apical and basal interfaces of the epithelium \cite{marciano2017holey}. Cancerous cells can disrupt normal tissue organisation in mammary glands by a process called luminal filling, associated with a loss of the luminal space \cite{muthuswamy2001erbb2, halaoui2017progressive}. Conversely, several cancerous cell types, when cultured {\it in vitro}, spontaneously self-organise in a spheroid surrounding a single or multiple lumens, e.g. cancerous breast cells \cite{Alladin2020} or colorectal cancer cells \cite{Ashley2013a}. Colorectal tumours also form multiple lumens surrounded by polarised epithelia {\it in vivo}; interestingly these spheroids can form with their apico-basal polarity oriented either towards or away from the central lumen \cite{Okuyama2016} (Fig.~\ref{Fig:introduction}e-f). 

Due to the fundamental physical mechanisms at play in lumen formation, these topics have attracted considerable attention within the biophysics community. In the next sections, we discuss some of the biophysical processes involved in lumen formation and interactions.








\section{Tissue pumping and the growth of single lumen}
\label{sec:single_lumen}
\begin{figure}[h!]
	\centering
	\includegraphics[width=0.75\linewidth]{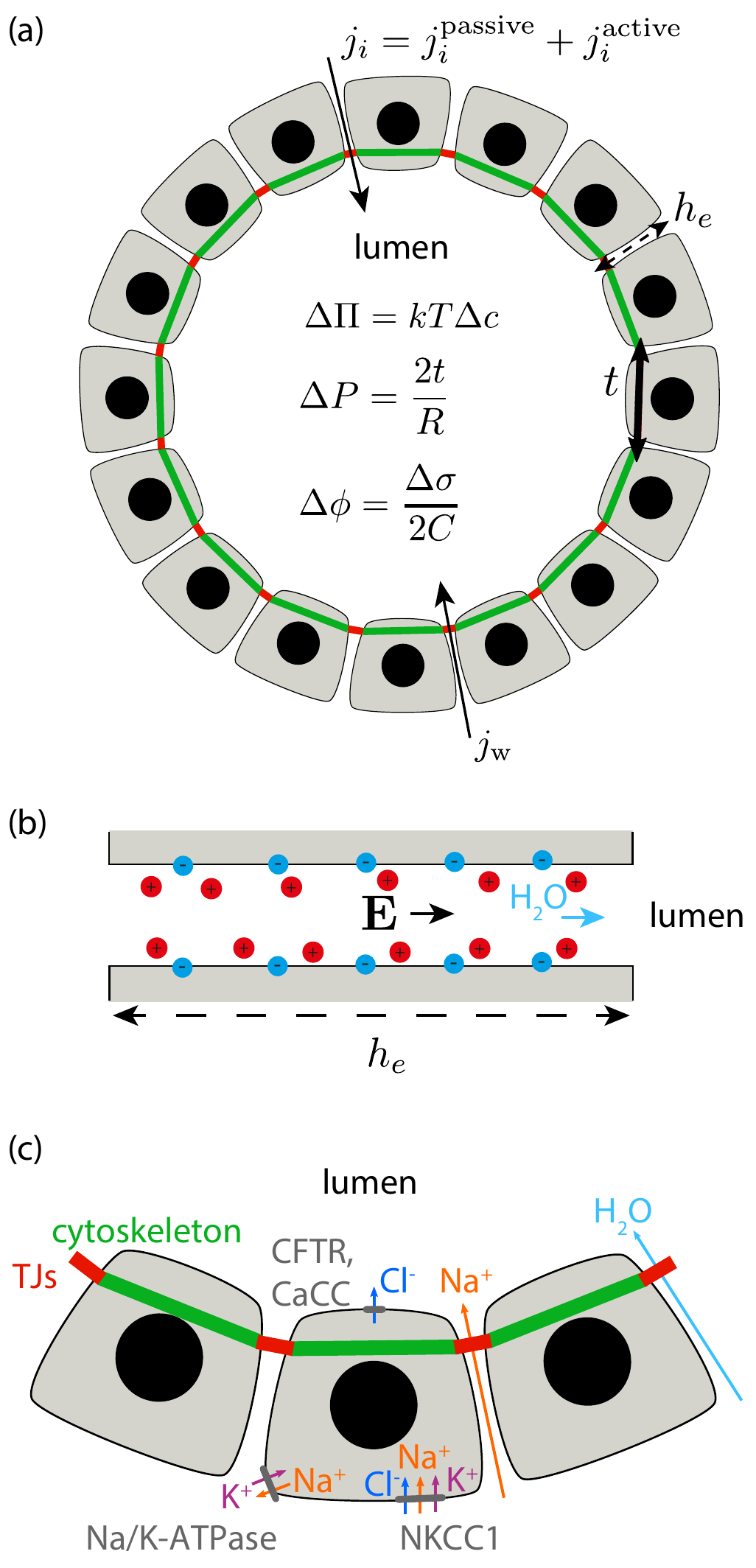}
	\caption{Mechanisms of lumen formation. (a) Water flows in the lumen due to a difference in osmotic pressure and hydrostatic pressures $\Delta\Pi-\Delta P$, and possibly through electro-osmotic effects. The hydrostatic pressure in a spherical lumen of radius $R$ is related to the tension $t$ in the monolayer through Laplace's law. The difference in osmotic pressure is mantained by the active transport of ions, which can also lead to a difference in electrostatic potential $\Delta \phi$. (b) Schematic of electro-osmotic effect: an extracellular fluid flow results from the electric field effect on free charges. (c) Mechanisms of active ion transport. The active transport of Na$^+$ in the basolateral side leads to a secondary active transport of Cl$^-$, which is exported to the lumen. In leaky epithelia, some ions, such as Na$^+$, flow through the intercellular space, in the paracellular domain, to balance the excess of negative charge in the lumen. }
	\label{Fig:pumping}
\end{figure}

\textbf{Irreversible thermodynamics of ion and water transport across an epithelium. }
To discuss the different mechanisms by which water can be transported across an epithelium, here we follow the framework of linear irreversible thermodynamics. In this framework, molecular fluxes are proportional to the chemical potential differences across the layer, with proportionality coefficients characterising the different couplings in the system \cite{kedem1963permeability, de2013non}. In the Box ``Passive flux of water and ions across a layer'' we give further details about the corresponding derivation, including the expression for the chemical potential of water and solutes. Assuming a dilute solution of $N$ ion types with concentrations $c_i$ and charges $q_i$, the {volumetric density} flux of water {towards the lumen} is given by the expression
\begin{equation}
\label{flux_water}
j_\text{w} = \lambda_\text{w} \left(\Delta \Pi - \Delta P\right) - \sum_{i=1}^N \lambda_{\text{w},i} \left(\Delta c_i+\frac{q_i \bar{c}_i}{k T} \Delta\phi\right)+j_w^{\rm active},
\end{equation}
where $\Delta c_i$ is the difference of concentration across the layer, i.e.~$\Delta c_i=c_i^l-c_i^0$ with $c_i^l$ the concentration in the lumen and $c_i^0$ the concentration outside, $\bar{c}_i=\left(c_i^l+c_i^0\right)/2$ is the average concentration,
$\Delta \Pi=kT \sum_{i=1}^N \Delta c_i$ is the osmotic pressure difference, $\Delta\phi$ the difference in electric potential, $\Delta P$ is the hydrostatic pressure difference, $\lambda_\text{w}$ is the permeability of the layer to water and $\lambda_{\text{w},i}$ is a cross-coupling coefficient characterising the flow of water that is driven by differences of solute chemical potential across the epithelium (Fig.~\ref{Fig:pumping}a). The flux $j_\text{w} $ is given in units of a permeation velocity. In the right-hand side of Eq.~\eqref{flux_water}, the first two terms correspond to passive water flows due to differences in its chemical potential and to dissipative couplings with other molecules in the solution.

In the last term, we have also included a possible active contribution to water pumping. 
We are not aware of a known active mechanism of cellular water transport. {We note however that apico-basal actomyosin cortical flows could in principle bring such an active contribution, due to friction forces between the paracellular fluid and the cortical cytoskeleton. Such friction forces have been proposed to allow for cellular swimming \cite{farutin2019crawling}, and friction forces between the actomyosin cortex and the cell cytoplasm is thought to be at the origin of the cytoplasmic flow in the C.elegans embryo, with a velocity $\sim 0.1$ \textmu m/s \cite{niwayama2011hydrodynamic}. This large value compared to reported luminal growth velocities \cite{Mosaliganti2019} suggest that such flows could contribute to water pumping, if they exist with the right magnitude and direction. }

The flux of solutes is given by
\begin{equation}
\label{flux_ions}
\begin{aligned}
j_i =& -\lambda_i \left(\Delta c_i+\frac{q_i \bar{c}_i }{k T}\Delta\phi \right) + \lambda_{i, \text{w}} \left(\Delta \Pi - \Delta P\right) \\
&- \sum_{j\neq i} \lambda_{i,j} \left(\Delta c_j+\frac{q_j \bar{c}_j}{k T} \Delta\phi\right)+j_i^{\rm active},
\end{aligned}
\end{equation}
where $\lambda_i$ is the permeability of the layer to solute $i$, and $\lambda_{i,\text{w}}$, $\lambda_{i,j}$ are cross-coupling coefficients characterising the flow of solute that is driven by differences of chemical potential of water and other solutes, across the epithelium. Here $j_i$ has the usual units of a flux density, i.e.~number of particles per unit area and time. As for the water flux, in the right-hand side of Eq.~\eqref{flux_ions}, the first three terms describe passive fluxes of solutes, while the last term describes active solute pumping by the epithelium.
The coefficients $\lambda_{\text{w}}, \lambda_i, \lambda_{\text{w},i}, \lambda_{i,j}$ form the matrix of phenomenological coefficients and encode all possible passive interactions between molecules across the layer. We now look at the interpretation of these parameters based on previous studies. 

\begin{figure*}
\begin{mdframed}
\textbf{Passive flux of water and ions across an epithelial layer}.  Here we discuss the equations determining the flux of water and ions across a layer following the framework of irreversible thermodynamics. We consider the free energy density of a solution of $N$ ion types in water
\begin{equation}
\label{eq:free_energy}
	\begin{aligned}
	f\left(c_\text{w},\{c_i\}_1^N\right) = & \mu^\text{ref}_\text{w} c_\text{w}+ kT c_{\rm w} \log\left(\frac{c_\text{w}}{c_\text{w}+\sum_{i=1}^N c_i}\right) +  \sum_{i=1}^N c_i \left[ \mu^\text{ref}_i +{kT} \log\left(\frac{c_i}{c_\text{w}+\sum_{i=1}^N c_i}\right)  +  q_i \Phi\right]\\
	&{+ P\left(c_\text{w} v_\text{w} + \sum_{i=1}^N c_i v_i-1\right)}~,
	\end{aligned}
\end{equation}
where $c_\text{w}$ is the concentration of water, $\mu^\text{ref}_\text{w}$ {a water} reference chemical potential, $c_i$ is the concentration of the $i$-th solute, $q_i$ its charge, $\mu_i^\text{ref}$ {a solute} reference chemical potential, $\Phi$ is the electric potential, {$k$ and $T$ are the Boltzmann constant and temperature}. The second and third terms in the right-hand side of Eq.~\eqref{eq:free_energy} are respectively the entropy density of mixing and the potential energy of charges in solution. $P$ is the pressure in the system{, and $v_{\text{w}}$ and $v_i$ the molecular volumes of water and osmolytes respectively. Here the pressure $P$ enforces an incompressibility condition $c_\text{w} v_\text{w} + \sum_{i=1}^N c_i v_i=1$}. The chemical potentials are given by
\begin{equation}
\begin{aligned}
\mu_\text{w} &= \frac{\partial f}{\partial c_{\rm w}} &\approx \mu_{\rm w}^\text{ref}+v_\text{w}\left( P - kT\sum_{i=1}^N c_i\right),\\
\end{aligned}
\end{equation}
\begin{equation}
\begin{aligned}
\mu_i &=  \frac{\partial f}{\partial c_i} 
\approx \mu_{i}^\text{ref}+kT \log( {v_w}c_i )+ q_i \Phi + P v_i,
\end{aligned}
\end{equation}
where in the last expressions we have written the leading order terms for a dilute, incompressible mixture in which $c_i/c_\text{w} \ll 1$ and $c_\text{w} v_\text{w} + \sum_{i=1}^N c_i v_i=1$. The passive flux density across the epithelial monolayer of water, {$\hat{j}_\text{w}$}, and ions $j_i$, will then follow the gradient of their chemical potentials. Following the framework of linear irreversible thermodynamics \cite{de2013non}, we have that
\begin{equation}
\begin{pmatrix}
\hat{j}_\text{w}\\
j_1\\
\vdots\\
j_N
\end{pmatrix} =- \begin{pmatrix}
\bar{\lambda}_\text{w} & \bar{\lambda}_{\text{w}1} & \cdots & \bar{\lambda}_{\text{w}N}\\
\bar{\lambda}_{\text{w}1} & \bar{\lambda}_1 & \cdots & \bar{\lambda}_{1N}\\
\vdots & \vdots & \ddots & \vdots\\
\bar{\lambda}_{\text{w}N} & \bar{\lambda}_{1N} & \cdots & \bar{\lambda}_N
\end{pmatrix}
\begin{pmatrix}
\Delta \mu_\text{w}\\
\Delta \mu_1\\
\vdots\\
\Delta \mu_N
\end{pmatrix},
\end{equation}
where the matrix is the symmetric and positive semidefinite matrix of Onsager coefficients. To get Eqs.~\eqref{flux_water} and \eqref{flux_ions}, we define the permeation velocity of water in terms of its flux density as $j_\text{w} = \hat{j}_\text{w} v_\text{w}$. For the chemical potential of ions, we expand {concentrations $c_i$ around a mean concentration $\bar{c}_i$}:
\begin{equation}
\label{eq:box_chem_pot}
\Delta\mu _i\approx  \left[kT \frac{\Delta c_i}{\bar{c}_i}+ q_i\Delta\phi  +\Delta P v_i \right],
\end{equation}
where, taking into account that $kT/v_i\sim 100$ {M}Pa, and since a typical excess pressure in the lumen is $\Delta P\sim 0.1$ kPa \cite{Latorre2018}, we assume that the last term can be neglected. Finally we define the parameters $\lambda_\text{w} = \bar{\lambda}_\text{w} v_\text{w}^2$, $\lambda_i = \bar{\lambda_i} k T /  \bar{c}_i$, $\lambda_{i,j} = \bar{\lambda}_{ij} k T / \bar{c}_j$, $\lambda_{\text{w},i} =\bar{\lambda}_{\text{w}i} k T v_\text{w} / \bar{c}_i$ and $\lambda_{i,\text{w}} = \bar{\lambda}_{\text{w}i} v_\text{w}$.
\end{mdframed}
\end{figure*}

{\textbf{Transepithelial water flux and pressure difference.}}
The epithelial permeability to water is characterised by $\lambda_\text{w}$ and encodes the effective permeability to both paracellular flows across the LIS and transcellular flows occurring across the lipid membrane or aquaporins. For MDCK cells $\lambda_\text{w}\sim (0.1-1) \cdot 10^{-7}$ \textmu m.s$^{-1}$.Pa$^{-1}$ \cite{Timbs1996-gt,Latorre2018}. This number is comparable to typical values of lipid membranes water permeabilities $\lambda_m \sim 0.7 \times10^{-7} $ \textmu m.s$^{-1}$.Pa$^{-1}$ (corresponding to a membrane permeability coefficient $P_f=10$ \textmu m.s$^{-1}$ \cite{olbrich2000water}; {this value depends on the aquaporin density in the membrane \cite{verkman2000structure}}). Higher values have been reported for other epithelia ($\sim 30-100$ times larger for the gallbladder and for the kidney proximal tubule \cite{fischbarg2010fluid}). During growth of the zebrafish inner ear, the water flux $j_\text{w}$ is of the order of $1-8$ \textmu m/h \cite{Mosaliganti2019}; corresponding to a driving pressure difference $\Delta \Pi - \Delta P \sim  3-200$ kPa with the values of $\lambda_{\rm w}$ {reported above for MDCK cells}, using Eq.~\eqref{flux_water} and neglecting cross-coupling coefficients. This relatively high pressure difference, compared to typical cytoskeletal pressures of $\sim 100$ Pa \cite{Salbreux2012}, is however consistent with physiological osmotic pressure differences: using Van't Hoff law, a difference of concentration of $1-100$ mM across the epithelium yields an osmotic pressure difference $\Delta \Pi\sim 2.6-260$ kPa. Direct pressure measurements of the intraluminal pressure in the growing zebrafish otic vesicle, using a sensor coupled to a glass needle penetrating into the vesicle, indicate lower values for the excess pressure $\Delta P$ of the order of $100-300$ Pa \cite{Mosaliganti2019}; similar values for $\Delta P$ are found in the blastocoel ($\Delta P \sim 300$ Pa) \cite{Dumortier2019}, in epithelial domes ($\Delta P \sim 100$ Pa) \cite{Latorre2018}, and in MDCK cysts ($\Delta P \sim 40$Pa \cite{narayanan2020osmotic} which was found to decrease to $\sim20$ Pa following inhibition of the CFTR channel). These observations suggest a mode of lumen growth which occurs in the regime $\Delta \Pi\gg \Delta P$. 

We also note that the difference of pressure $\Delta P$ in Eq.~\eqref{flux_water} is not necessarily constant, and in particular dissipative processes associated to tissue deformation might also contribute to $\Delta P$ and further slow down, or entirely control, the speed of lumen expansion. Considering the epithelium as a purely fluid layer with two-dimensional viscosity $\eta$, {Laplace's law leads to $\Delta P = 2t/R=(4\eta/R^2) \dot{R}$ with the tension due to the epithelial deformation $t=2\eta \dot{R}/R$. For a spherical lumen, using Eq.~\eqref{flux_water} giving here $\dot{R} = \lambda_\text{w} \left[\Delta \Pi - \Delta P\right]$, one would obtain $(1+ 4 \lambda_\text{w} \eta/R^2)\dot{R} = \lambda_\text{w} \Delta \Pi $. Thus, the relative role of tissue viscosity and fluid permeation in controlling the speed of lumen expansion is set by the dimensionless ratio $4\lambda_{\rm w} \eta/R^2$.} With a typical 3D tissue viscosity $\eta_{\rm 3D} \sim 10^5-10^6$Pa.s \cite{marmottant2009role, guevorkian2010aspiration}, $\eta\sim h \eta_{\rm 3D}$ with $h\sim 10$ \textmu m the tissue thickness and $R\sim 100$ \textmu m, one obtains $4\lambda_{\rm w} \eta/R^2\sim 4 \times 10^{-6}-10^{-4}$, indicating that water permeation {can be} the dominating dissipative process. {More generally the state of the tissue can change as the lumen grows, for instance due to cellular deformations leading to elastic stresses \cite{Mosaliganti2019}, cell division and death, or cell volume change \cite{hoijman2015mitotic}, leading to changes in tissue tension and, as a result of the law of Laplace, excess hydrostatic pressure within the lumen.}

\textbf{Ion permeability of the epithelium. }
The epithelial ion permeability, relating the ion flux density across the layer to a gradient in its chemical potential, is characterised by $\lambda_i$ and, as well as $\lambda_\text{w}$, it effectively encodes the permeability for both paracellular and transcellular fluxes, including the effect of ion channels, pores, carriers and symporters. This permeability however excludes the effect of pumps: pumps are indeed actively transporting ions by consuming a source of chemical energy such as ATP; we will discuss them later. The paracellular ion  permeability depends on the permeability of tight junctions, some of which can be cation/anion selective \cite{Krug2014-ef}. For instance, tight MDCK monolayers, which are devoid of claudin-2 tight junction proteins that facilitate the permeation of small cations such as Na$^+$ or K$^+$, develop very large transepithelial potentials and very large electric resistance $\sim 10$ k\textOmega cm$^2$. When transfected with claudin-2-cDNA, the resistance lowers to 150-500 \textOmega cm$^2$ \cite{Furuse2001-jo}. Measures of $\lambda_i$ are scarce in the literature but an order of magnitude can be obtained from measurements of epithelial electric resistance \cite{Dasgupta2018}. Given that $q_i j_i$ is an electric current, $\omega_i=kT/(\lambda_i \bar{c}_i q_i^2)$ is an electric resistance times unit area, which for epithelial monolayers is typically in the range $10-{10000}$ \textOmega.cm$^{2}$ \cite{Sackin2013-fw}. Using typical ion concentrations of $100$ mM \cite{butt2013physics} and $q_i=e$ the electron charge, we can obtain an estimate for the permeability $\lambda_i=3\times {10^{-3}}-3$ \textmu m/s, in agreement with the permeability of Na$^+$ in MDCK monolayers $\lambda_i=1.7\times 10^{-1}$ \textmu m/s \cite{Fernandez-Castelo1985-pi}.

 If one imposes a difference in solute concentration across an epithelium surrounding a lumen of radius $R$,  both a water flux and a solute flux results from this difference. {As a result of the solute flux, the solute concentration difference decays on a time scale $\sim R/(3 \lambda_i)$. With the estimate for ion permeability above and $R=100$ \textmu m, this gives equilibration occurring on a time between $3$ s and $3$ hours. Water fluxes driven by difference of osmotic pressures are sustained on longer time scales if active solute fluxes constantly compensate for this solute leakage, as we discuss below.}
 

In leaky epithelia, with a low electric resistance, the difference of electric potential across the epithelium $\Delta \phi$  is small compared to tight epithelia ($\Delta \phi \sim\pm 50$ mV) \cite{Reuss2008-nw, Sackin2013-fw}. 
 Denoting $C$ the capacitance per unit area of the epithelium (with order of magnitude $C\sim 10^{-2}$ pF/\textmu m$^2$ \cite{Wegener1996-zc}), the transepithelial potential satisfies $\Delta \phi = \Delta\sigma /({2}C)$, where $\Delta\sigma$ is the difference of charge density across the epithelium. Assuming that ions equilibrate quickly in the lumen, excess freely-diffusing charges in the lumen will be located in a thin layer covering the epithelium with a typical thickness given by the Debye-H\"uckel length scale $\ell_\text{DH}$ (see Box ``Interaction of charged surfaces''); giving rise to the charge difference $\Delta\sigma$. Assuming also that the solution in the basal side is electrically neutral, $\Delta\sigma = V \sum_i q_i^l c_i^l/A$ with $V$ and $A$ the volume and surface area of the lumen. 

\textbf{Ion and water interactions across the epithelium. }
The cross-coupling coefficients $\lambda_{\text{w},i}$, $\lambda_{i,\text{w}}$ and $\lambda_{i,j}$, which satisfy $\lambda_{i,\text{w}} = \lambda_{\text{w},i} \bar{c}_i/kT$ and $\lambda_{i,j} \bar{c}_j = \lambda_{j,i}\bar{c}_i$ due to the underlying Onsager reciprocal relations, stem from passive interactions between water and ions and different ions across the monolayer. 

Cross-couplings between ion transport characterised by the coefficients $\lambda_{i,j}$ can be attributed to cotransporters which transport several species simultaneously across the cell membrane. Symporters such as NKCC1 can use favorable differences in chemical potential of some of the transported species (e.g. Na$^+$ in the case of NKCC1) to transport other molecules against their electrochemical potential (e.g. Cl$^-$ and K$^+$ in the case of NKCC1 \cite{Barrett2000-gt}). The coefficients $\lambda_{i,j}$ therefore effectively appear in models for ion transport across the monolayer which incorporate fluxes due to symporters, whose rate of transport is driven by concentration differences of transported ions \cite{benjamin1997quantitative,Gin2007-oc}.

The coefficients $\lambda_{\text{w},i}$ give rise to the osmotic Staverman's reflection coefficients, $r_i = 1-\lambda_{\text{w},i}/(kT\lambda_{\text{w}})$ \cite{staverman1951theory, kedem1963permeability}. For fully impermeable species which do not cross the epithelium, $\lambda_i = \lambda_{i, \text{w}}=\lambda_{\text{w},i}=0$ and the corresponding reflection coefficient $r_i=1$. What is the possible origin of the cross coupling coefficients  $\lambda_{i, \text{w}}$, $\lambda_{\text{w}, i}$, in epithelial monolayers? In \emph{Xenopus} oocytes the Na$^+$/glucose symporter has been shown to cotransport water \cite{Loo1996-tl} which leads to a nonzero $\lambda_{\text{w},\text{Na}^+}$ coefficient. Water can also in principle be dragged due to paracellular flows of solutes across paracellular junctions \cite{Ussing1989-hd}. In the case of an uneven coupling between water with anions and cations, for instance because of symporters cotransporting water, or because some ionic species bind more strongly to the cell surface than others, the theory predicts an electro-osmotic coupling, i.e.~a water flux $j_{\rm w}$ driven by a transepithelial potential $\Delta \phi$, in the absence of an osmotic pressure difference \cite{fischbarg2010fluid}. In the corneal endothelium, a coupling between water flows and ionic current has been measured with a magnitude $2.4$ \textmu m.cm$^2$/(h \textmu Amp) \cite{Sanchez2002-vh}. Given that the electric resistance of the endothelium $\omega=20$ \textOmega.cm$^2$, this leads to a measure for $\lambda_{\text{w},i} q_i \bar{c}_i/kT$ that can be used to obtain e.g. $\lambda_{\text{w},\text{Na}^+}\sim 1.4\cdot 10^{-8}$ \textmu m$^4$/s. 

For comparison, we can discuss the electro-osmotic flux that arises in the intercellular space as a response to an electric field, triggering the motion of the layer of freely diffusing ions located near the charged membrane  \cite{sarkar2019field} (Fig.~\ref{Fig:pumping}b). We assume that an electric field $-\Delta \phi/h_e$ with $h_e$ the tissue apico-basal height, oriented along the apico-basal axis, is present in the intercellular space. We also assume that cell membranes lining the intercellular space are charged, with the charge characterised by the cell membrane zeta-potential $\zeta$; and that Na$^+$ is providing the counterions. The electric field then gives rise to a plug flow in the intercellular space, whose velocity is given by the Helmholtz-Smoluchowski formula, $v\sim   \epsilon \zeta/(\eta_{\rm w} h_e)\Delta \phi$, with $\eta_\text{w}$ the water viscosity \cite{butt2013physics}. The corresponding flux of water across the epithelium is $j_\text{w}= \phi_{\rm LIS} v$ with $\phi_{\rm LIS}$ the volumic fraction of intercellular space in the tissue. Identification with Eq.~\eqref{flux_water} then gives $\lambda_{\text{w},\text{Na}^+}\sim -2\ell_{\rm DH}^2 e \zeta \phi_{\rm LIS}/(\eta_{\rm w} h_e)$. With $\eta_{\rm w}=10^{-3}$Pa.s, $\zeta=-30$ mV \cite{bondar2012monitoring}, $h_e=10$ \textmu m, $\ell_{\rm DH}\simeq 1$ nm and $\phi_{\rm LIS}=10^{-2}$, one finds indeed a comparable value of $\lambda_{\text{w},\text{Na}^+}\sim 10^{-8}$ \textmu m$^4$/s. With these numbers, a difference of electric potential of $-1$ mV can elicit a water flow with velocity $0.02$ \textmu m/s \cite{Mosaliganti2019}; a sufficiently large number compared to reported flows in lumen formation of $\sim$ \textmu m/h for this effect to play a role in practice. Whether or not these cross-coupling effects contribute to water pumping might depend on the type of epithelium \cite{fischbarg2010fluid}.



\textbf{Active solute pumping. }
The differences in ionic concentrations across the epithelium that drive passive water flows given by the mechanisms described in the previous paragraphs are set by the effect of ionic pumps, which generate an active flux density $j_i^\text{active}$. In the steady state, the total flux $j_i=j_i^\text{passive}+j_i^\text{active}=0$. 
Although numerous pumping mechanisms in epithelial cells have been reported, see \cite{Reuss2008-nw, Frizzell2012-kx} for an extensive review, here we describe the main processes {that have been proposed to lead to} active transport of ions {towards the lumen \cite{Frizzell2012-kx, Navis2015}} (Fig.~\ref{Fig:pumping}c). 
By hydrolysing ATP, the Na$^+$/K$^+$-ATPase pump transports three Na$^+$ from the cytoplasm to the interstitial fluid and two K$^+$ in the opposite direction per cycle with a {pumping rate} of the order of $10^2$ s$^{-1}$ \cite{Reuss2008-nw}. 
The activity of the Na$^+$/K$^+$-ATPase pump, as well as its basolateral distribution following the establishment of the cell apico-basal polarity, {has been reported to be} required for lumen formation, e.g.~inhibition of the Na$^+$/K$^+$-ATPase pump by ouabain blocks lumen formation \cite{Bagnat2007}. 
The Na$^+$/K$^+$-ATPase pump generates an electrochemical driving force for the entry of Na$^+$ into the cell, which through symporters such as NKCC1 leads to the import of other ions such as Cl$^-$ \cite{Barrett2000-gt}. Chloride is then exported to the lumen through channels such as CaCC or CFTR.
Inhibition/activation of these channels leads to decreased/increased lumen size \cite{li2004relationship,Bagnat2007}.
Altogether, this leads to ionic flux of Cl$^-$ across the monolayer which is indirectly actively driven. 
In turn, this generates an electrochemical gradient across the epithelium which draws a passive paracellular flow of Na$^+$, i.e. across the intercellular space, into the lumen. {As another possible active mechanism of osmolyte transport to the lumen, exocytic vesicles have been observed concomitantly with lumen initiation in several systems \cite{Bryant2010, Ryan2019}.}

\textbf{A simple model for the growth of a spherical lumen.}
We now discuss how the previous ingredients can be used to derive a simple picture for the growth of a spherical lumen. For simplicity, we do not take into account here the different passive cross-coupling terms between water flux and ion chemical potential difference, conversely cross-couplings between ion fluxes and water chemical potential difference, {and active water transport}.
For a spherical lumen of radius $R$, {conservation of lumen volume and number of molecules and }Eqs.~\eqref{flux_water} and \eqref{flux_ions} give for the rate of change of the radius $R$ and lumen concentration $c_i$,
\begin{equation}
\begin{aligned}
\label{eq:dynamic_lumen}
\dot{R} &= \lambda_\text{w} \left[\Delta \Pi - \Delta P\right],\\
\frac{R}{3}  \dot{c}_i^{\rm l}&=-\dot{R} c_i^{\rm l}  -\lambda_i \left[\Delta c_i+\frac{q_i \bar{c}_i }{k T}\Delta\phi  \right]+ j^\text{active}_i,
\end{aligned}
\end{equation}
where $\Delta \phi = R/({6}C) \sum_i q_i \Delta c_i$, $\Delta \Pi = kT \sum_i \Delta c_i$ and the difference of hydraulic pressure satisfies Laplace's law $\Delta P = 2t/R$ with $t$ the surface tension in the monolayer. {The surface tension of the monolayer might depend on $R$ and $\dot{R}$ through the constitutive law of the monolayer}. In Eq.~\eqref{eq:dynamic_lumen} we have made the approximation that the volume change in the lumen is entirely due to the water flux across the epithelium. For simplicity, we can consider $j^\text{active}_i$ to be constant. Assuming a density of $10^3-10^4$ \textmu m$^{-2}$ Na$^+$/K$^+$-ATPase pumps \cite{Ewart1995-hy} with a {pumping rate} of $10^2$ s$^{-1}$ \cite{Reuss2008-nw}, $j_i^\text{active}\sim 10^5-10^6$ \textmu m$^{-2}$ s$^{-1}$, whose range is consistent with pumping rates measured across MDCK monolayers \cite{Cereijido1981-bz, Simmons1981-ok, Latorre2018}, as well as net Na$^+$ fluxes measured across the rabbit corneal endothelium \cite{lim1982analysis}.  Similar equations have been used in recent studies \cite{Gin2010-wf,Ruiz-Herrero2017,Mosaliganti2019,Chan2019}, although the transepithelial potential is not always explicitely considered. 

To clarify the discussion, we consider a simple case of a pump-leak mechanism where an anion with charge $-e$ and concentration $c_-$ is pumped towards the lumen with an active flux density $j^{\rm active}$, while a corresponding cation with charge $+e$ with concentration $c_+$ diffuses through the epithelium passively. The external concentrations are also taken to be fixed, $c^{\rm o}_+=c^{\rm o}_-=c_0$; and the permeability of the epithelium to both ions taken to be equal, $\lambda=\lambda_+=\lambda_-$. At low epithelial capacitance $C$ and for constant lumen radius $R$, the steady-state concentration differences are $\Delta c_+=\Delta c_-=j^{\rm active}/(2\lambda)$ and the transepithelial potential $\Delta \phi =-(\Delta c_+/\bar{c}) (kT/e)$ ($\bar{c}\simeq c_0$ here as we assume small concentration differences). In this expression, $kT/e$ is a characteristic electric potential which is about $27$ mV, an order of magnitude comparable to measured transepithelial potentials. With the ranges discussed above $j^{\rm active} =10^5-10^6$ \textmu m$^{-2}.{\rm s}^{-1}$ and $\lambda=3\times 10^{-2}-3$ \textmu m.s$^{-1}$, this gives $\Delta c$ ranging from $0.03$ mM to $30$ mM -  values towards the end of this range are consistent with physiological concentration of ions of $\sim 100$ mM \cite{butt2013physics}.  

The equilibrium radius for the lumen is then given by the balance of hydrostatic and osmotic pressure {($\Delta P = \Delta \Pi$)}, giving {$2t/R = \Delta P = \Delta \Pi = kT\left(\Delta c_++\Delta c_-\right) =  k Tj^{\rm active}/\lambda$}.  Depending on the constitutive law for the monolayer tension and dependency of the active flux density $j^{\rm active}$ on the lumen size, one expects the corresponding equilibrium lumen size given by the radius $2t\lambda/(k Tj^{\rm active})$ to be stable or unstable: e.g.~if the epithelial tension $t$ is independent of strain and the active flux density $j^{\rm active}$ is constant, the equilibrium state is unstable as an increase of the radius leads to a decrease in the hydrostatic pressure and further flow towards the lumen; this point is further discussed in the next section. Elastic stresses in the epithelium would generally tend to stabilise the lumen  \cite{Mosaliganti2019}. Indeed, considering an elastic epithelium with surface tension given by $t=t_0+2K \delta R/R_0$ with $t_0$ and $R_0$ the steady-state epithelial tension and radius, $\delta R$ a perturbation of the radius and $K$ the epithelial area elastic modulus, the pressure change in the lumen due to a change in epithelium radius is given by $\delta P= 2[2K-t_0]\delta R/R_0^2${, which, for $K>t_0/2$, appears as a stabilisation term for lumen growth in Eq.~\eqref{eq:dynamic_lumen}.} Physically, an increase in lumen radius leads to a positive tension in the epithelium due to elastic stretching, increasing the excess hydrostatic pressure in the lumen and favouring fluid expulsion.

The steady-state relations obtained above indicate that lumens do not need to be perfectly sealed to grow: the balance between the active pumping density flux $j^{\rm active}$ and leakage with rate $\lambda$ can set up a difference of concentration gradient allowing for lumen expansion. Indeed, pumping of ions within a forming cavity leads to a positive osmotic pressure which favours the growth of the cavity and counteracts its tendency to close due to the cortical surface tension \cite{Dasgupta2018}. Leakage of ions is opposing the growth of the cavity, due to the reduction of the osmotic pressure associated to ion loss. A dynamic balance between these effects implies that for any value of the leakage magnitude, a threshold rate of pumping can compensate and ensure that the cavity is growing  \cite{Dasgupta2018}.
 
In a dynamically growing lumen, the luminal concentrations are also changing by dilution, due to incoming water fluxes. Keeping explicit forms for the flux densities $j_i$ and $j_\text{w}$, the change of concentration in the lumen can indeed be written:
\begin{align}
\frac{R}{3} \dot{c}_i^{\rm l} = j_i - j_\text{w} c_i^{\rm l}~,
\end{align}
such that a constant luminal concentration $c_i^{\rm l}$ can be maintained during lumen expansion if the water and solute fluxes balance according to $j_i=c_i^{\rm l} j_\text{w}$ \cite{Mosaliganti2019}. If the role of the hydrostatic pressure difference $\Delta P$ can be neglected compared to the osmotic pressure difference $\Delta\Pi$ and the influx of solute $j_i$ is constant, the model then predicts  the relaxation to a constant solute concentration in the lumen, $c_i^{\rm l}=j_i/j_\text{w}$ and convergence to a linear radius increase $dR/dt=j_\text{w}$.

\textbf{Flexoelectricity. }
In this discussion we have incorporated a simple description of electric effects. A recent theoretical work has studied a possible role in lumen formation for flexoelectricity, the establishment of an electric current in the tissue that is sensitive to tissue curvature \cite{Duclut2019a}. Flexoelectricity can lead to an effective negative surface tension of the inner interface of the tissue facing the lumen, generally favoring growth and allowing for lumen nucleation, even when the osmotic pressure difference is unfavorable to lumen expansion.

\section{Lumen interactions: Physics of hydraulic flows in tissues}
\label{sec:lumen_interactions}
\begin{figure}[h!]
	\centering
	\includegraphics[width=0.9\columnwidth]{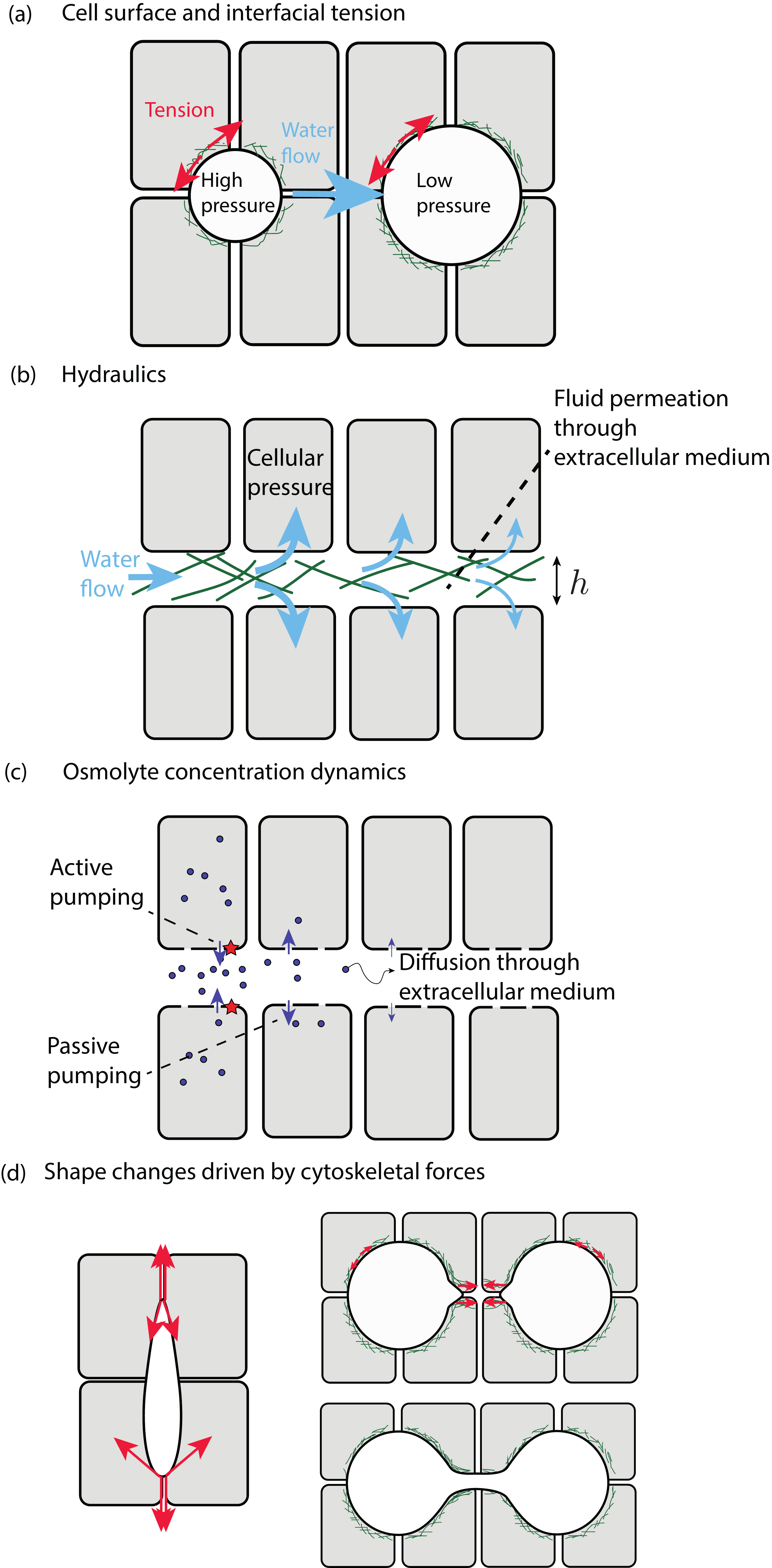}
	\caption{Interactions of several lumens. (a) Imbalance of lumen hydrostatic pressure can drive fluid flows. In the case of spherical lumens under constant surface tension, water flows from smaller towards larger lumen. (b) Fluid flows and pressure gradients depend on fluid permeation through the extracellular space and through cell membranes. (c) Gradient of extracellular osmolyte concentration depend on their diffusion through the interstitial space, as well as active and passive pumping through cellular membranes. (d) Lumen shapes can depend on gradients of cortical tensions \cite{Li2016}. Lumen coalescence can be driven by cellular contractions and rearrangements. }
	\label{Fig:lumen_interactions}
\end{figure}

\textbf{Lumen size instabilities. }
We now discuss interactions of several lumens in biological tissues. 
As discussed in the introduction, large lumen formation can occur through intermediate steps involving the growth and fusion of multiple smaller micro-lumens. In recent years several key points on the physics of lumen communication have been made. If lumens are enclosed by cellular interfaces under tension due to the action of the acto-myosin cytoskeleton  \cite{Salbreux2012}, this results in an excess hydrostatic pressure in the lumen given by the law of Laplace, $\Delta P=2 t/R$, with $t$ the epithelial surface tension and $R$ the radius of curvature. Such configuration is fundamentally unstable if the tension $t$ is constant (Fig.~\ref{Fig:lumen_interactions}a): indeed a reduction of the radius of curvature $R$ leads to a higher hydrostatic pressure difference $\Delta P$, which favours expulsion of the fluid out of the cavity and further reduction of the radius $R$. If the osmotic pressure difference between the lumen and the extracellular space, $\Delta \Pi$, is constant, a critical radius $R^*=2t/\Delta\Pi$ determines whether the lumen collapses (for $R<R^*$) or expands indefinitely (for $R>R^*$). By contrast, if the total osmolyte number is fixed inside the lumen, osmotic pressure provides a stabilising effect as constriction of the lumen leads to an increase in osmolyte concentration, and an increase in the lumen osmotic pressure difference $\Delta \Pi$, {triggering a water flow} which restores the lumen initial size.

Because of this fundamental instability, when several spherical cavities subjected to equal surface tension are brought into contact, a flow arises directed from smaller cavities towards larger cavities (Fig.~\ref{Fig:lumen_interactions}a). In binary mixtures that phase separate, a similar physical instability gives rise to Ostwald ripening. In this dynamic process, small droplets of one of the mixture component progressively coarsen through a diffusive flux which favours the growth of bigger droplets at the expense of smaller droplets \cite{yao1993theory}. As droplets coarsen,  the interfacial energy of the mixture is reduced, such that the end result, in a finite system, is a single droplet rich in one component, surrounded by a medium rich in the second component. Similarly, one would expect such a process of lumen coarsening in a tissue to eventually result in a single lumen. This is indeed the dynamics that has been experimentally observed in mouse blastocysts \cite{Dumortier2019}. In line with the picture of water flowing from microlumens of high pressure to microlumens of lower pressures, the position of the emerging single lumen can be biased experimentally by mixing  cells with higher and lower surface tension, through knock out of Myh9 \cite{Dumortier2019}; as expected the cells surrounding the winning lumen are more likely to have a lower tension. A characteristic scaling law of growth controls Ostwald ripening, such that the mean droplet radius $R$ increases with time {$\tau$} as {$R\sim \tau^{\frac{1}{3}}$} for diffusion-limited growth \cite{marqusee1983kinetics}. It is unclear if a similar scaling relation can also be determined in the process of lumen fusion and coarsening; {notably the physics of matter exchange might differ between classical Ostwald ripening and the coarsening of biological lumens \cite{Verge-Serandour2021}.}

\textbf{Extracellular water flows. }
How does water flow in response to gradients of hydrostatic pressure? It seems reasonable to assume that water flows in the extracellular space of a tissue according to Darcy's law, such that a gradient of hydrostatic pressure in the extracellular fluid, $P^f$, drives a flow of the extracellular fluid with average velocity $\mathbf{v}^f$, according to the relation \cite{Ranft2012}:
\begin{align}
\label{eq:Darcyslaw}
\boldsymbol{\nabla} P^f + \kappa(\mathbf{v}^f-\mathbf{v}^c)=0,
\end{align}
where $\mathbf{v}^c$ is the coarse-grained cell velocity in the tissue, and $\kappa$ is a {coefficient of hydraulic resistance}. The Darcy equation above describes flows in porous media; here the coefficient of {hydraulic resistance} $\kappa$ is associated to flow of the interstitial fluid through the extracellular matrix or through intercellular spaces in the tissue. If the extracellular fluid is flowing with viscosity $\eta_{\rm w}$ {in extracellular channels with volumetric density $\phi$}, and each cellular channel is filled with a filamentous mesh with filaments separated by a distance $\delta$,  the {hydraulic resistance } $\kappa\sim  32 \eta_{\rm w}/({\phi}\delta^2)$ (\cite{de1979scaling}, we {take into account here} a numerical prefactor here corresponding to Poiseuille flow in a tube of radius $\delta/2$). 
The typical time scale for pressure equilibration between lumens is then $\tau\sim R^2 \ell \kappa/ (2t) \sim16 R^2 \ell \eta_{\rm w}/ (t  \phi \delta^2)$; with $\ell$ a characteristic distance between lumens. With the viscosity of water $\eta_{\rm w}=10^{-3}$Pa.s, $t\sim 100$ pN.\textmu m$^{-1}$\cite{Salbreux2012}, $\delta\sim50$nm \cite{yurchenco1987basement}, and $R=\ell=20$ \textmu m, this gives $\tau>\sim 500$ s {using $\phi\leq 1$}. This timescale is roughly in line with the duration of lumen coarsening observed in the formation of the mouse blastocoel \cite{Dumortier2019}. We note however that the estimate for this timescale is highly sensitive to the actual value of the extracellular mesh size or gap size between cells, {and that non-filamentous extracellular structures, such as linker proteins between cells, could also limit extracellular flows}. {In the limit where the extracellular space is devoid of extracellular matrix and flows occur in intercellular channels of size $\delta_c$ separating cells with size $\ell_c$, $\phi\sim \delta_c^2/\ell_c^2$ and  $\tau\sim 16 R^2 \ell \eta_{\rm w} \ell_c^2/ (t  \delta_c^4)$. In the \emph{Xenopus}  embryonic ectoderm, interstitial gaps at tricellular junctions have been measured to have a characteristic size $\delta_c \sim 1-4$ \textmu m with a characteristic cell size $12$ \textmu m \cite{Barua2017}; resulting in faster equilibration time scale  $\tau\sim 1-184$s.} In addition, other dissipative processes, for instance associated to cellular deformations and rearrangements, might also limit the deformation of lumens in morphogenetic processes. 

This analysis of flow between lumens is modified if one takes into account that water can also flow across the cell membrane (\cite{Dasgupta2018}, Fig.~\ref{Fig:lumen_interactions}b). The competing processes of flow through the intercellular space and through the cell membrane give rise to a characteristic screening length, 
\begin{align}
\xi_\text{w} = \sqrt{\frac{{\delta_c}}{ 4\lambda_m {\phi} \kappa}}~,
\end{align}
with $\delta_c$ the width of the extracellular space, and $\lambda_m$ the membrane permeability to water. {This length scale arises by combining Darcy's law, Eq.~\eqref{eq:Darcyslaw}, with an equation for the conservation of matter, $\partial_{\alpha} v_{\alpha}^f  =-4 \lambda_m \phi/\delta_c (P^f-P^c)$, which takes into account losses of fluid through the cell membrane, here assuming cylindrical channels with diameter $\delta_c$ and a cellular pressure $P_c$.}
{If one considers a situation with fixed cellular pressure,} above this length-scale, extracellular flows are screened by absorption of the fluid into the surrounding cells. What is the order of magnitude of this length scale? With $\delta_c={1-4}$ \textmu m and $\lambda_m \sim 0.7 \times10^{-7} $ \textmu m.s$^{-1}$.Pa$^{-1}$ (discussed in section \ref{sec:single_lumen}), one finds $\xi=0.5-1$ mm.  This large length scale ($>\sim 50$ cell lengths, taking a typical cell diameter of $10$ \textmu m) indicates that hydraulic communication between lumens {through the extracellular space}  can be relevant to their dynamics. As noted above, the permeation constant $\kappa$ however varies strongly with the {extracellular matrix} mesh size $\delta$, and the length scale {$\delta_c$} for the dimension of the intercellular space is also likely to vary significantly, so that the precise value of the length scale $\xi_\text{w}$ could vary between experimental systems. {Also, this discussion assumes a constant cellular pressure and therefore does not take into account the possible role of transcellular flows.}

\textbf{Extracellular osmolyte diffusion. }
What is the dynamics of osmolytes (Fig.~\ref{Fig:lumen_interactions}c)? Osmolytes passively flow in the extracellular space and a diffusion flux should occur to balance concentration in different microlumens. Here as well such diffusive fluxes are screened by exchange with the intracellular space, through pumping of osmolytes across the cell membranes. The {corresponding screening} length scale can be written \cite{Dasgupta2018}
\begin{align}
\label{eq:osmolyte_length_scale}
\xi_i = \sqrt{\frac{D {\delta_c}}{ \lambda_{i}}}~,
\end{align}
with $D$ the diffusion constant of osmolytes in the extracellular medium, and $\lambda_{\rm osm}$ the membrane osmolyte permeability. A typical diffusion constant for ions in a tissue is $D\sim 500$ \textmu m$^2$.s$^{-1}$ \cite{suenson1974diffusion}. Taking {again } $\delta_c=1-4$ \textmu m and $\lambda_{i} =3\times 10^{-2}-3$ \textmu m/s, one finds $\xi_i\sim 12-240$\textmu m. This length scale arises from a competition between diffusion in the extracellular space and reabsorption in the cell. In a thick epithelium, this length scale could limit paracellular ion exchange across the epithelium, as ions would diffuse into the cells before crossing the epithelium. This length scale also determines, in principle, whether neighbouring microlumens share solutes that are pumped towards the lumens.

The length-scale involved in Eq.~\eqref{eq:osmolyte_length_scale} arises from a similar competition of processes to the diffusion-degradation model involved in morphogen gradient formation and spread, determining a patterning length scale $\sqrt{D/k}$ with $D$ the morphogen diffusion constant and $k$ an effective degradation rate \cite{Kicheva2007}. In Eq.~\eqref{eq:osmolyte_length_scale}, osmolytes are assumed to flow passively through the membrane instead of binding to cell membrane receptors and being internalised. In both cases, this length-scale determine the range of cell-cell communication occuring through diffusible molecules. The presence of the {$\delta_c$} factor in Eq.~\eqref{eq:osmolyte_length_scale} indicates that the length scale increases with the size of the intercellular space. Interestingly microlumens forming in the zebrafish lateral line primordium have been called ``luminal hubs'', as cells sharing a common lumen exhibits a coordinated behaviour as a result of FGF concentrating specifically in the shared lumen \cite{durdu2014luminal}.

\textbf{Nonspherical lumens.}
The shape of forming lumens is not necessarily spherical (Fig.~\ref{Fig:lumen_interactions}d): indeed due to the Young-Dupré force balance equation at the junctions between several cells, the interface surfaces can in principle deform away from a spherical shape to establish force balance at the cellular junctions. In liver cells cultured {\it in vitro}, early lumens open preferentially towards the tissue interface with the free medium, away from the extracellular matrix \cite{Li2016}. This asymmetric lumen growth process has been suggested to arise from differences in junctional tensions across cellular junctions. {In the zebrafish otic vesicle, the lumen grows anisotropically \cite{hoijman2015mitotic}. Such anisotropic shape is initiated early before the lumen starts to expand through the formation of an anisotropic apical surface, and is maintained through differential behaviour of the cells facing the lumens, notably through inhomogeneous epithelial thinning \cite{hoijman2015mitotic}. In MDCK cysts, the lumen solidity, a geometric measure of shape regularity that compares a given shape to its convex hull, changes with its size, with smaller lumens exhibiting more irregular shapes \cite{vasquez2020physical}. In general, one expects that lumen with volume close to the cell size are more irregular, due to the role of single cell mechanical effects playing a more visible role at this scale.} 

Differences in interfacial tension could also in principle drive lumen fusion, bringing lumens into contact for them to undergo coalescence by contact (Fig.~\ref{Fig:lumen_interactions}d). During zebrafish gut development, the intestinal tube forms in several stages, starting with the initiation of multiple small lumens within a solid rod of endodermal cells, followed by lumen fusion and resolution \cite{Alvers2014}. This fusion process could occur through de-adhesion of cellular interfaces in the bridge connecting adjacent lumens, of through contraction of the cellular interfaces participating to the bridge \cite{Alvers2014}. More generally, coalescence by direct contact (Fig.~\ref{Fig:lumen_interactions}d) or ``ripening'' based on fluid flows between lumens (Fig.~\ref{Fig:lumen_interactions}a) are two processes that could be responsible for lumen fusion, separately or in combination. 

\section{Conclusion}

The physics of lumen formation involves multiple length scales and a complex interplay of hydraulic flows, cell mechanics, and electric and electro-osmotic effects. This topic has attracted significant interest in recent years, as it is becoming evident that the dynamics and mechanics of lumen formation plays a key role in morphogenetic processes during development. Stem cells aggregates have the ability to spontaneous form a lumen, indicating that lumen formation is a fundamental self-organising ability of biological tissues. How this self-organisation occurs is a key question at the interface of physics and biology. The interplay of lumen formation with cellular forces and pumping, but also with cell polarity, morphogen gradient and cell signalling, and electric forces and currents leads to a large range of situations and rich physical behaviours.

\section*{Acknowledgement}
{GS thanks Jacques Prost for discussions and helpful feedback on the manuscript.}
ATS, MKW, and GS acknowledge support from the Francis Crick Institute, which receives its core funding from Cancer Research UK (FC001317), the UK Medical Research Council (FC001317), and the Wellcome Trust (FC001317).

\section*{Competing interests}
The authors have no competing interests to declare.
\bibliographystyle{ieeetr}
\bibliography{Review.bib} 
\end{document}